\documentstyle[12pt]{article}
\begin{document}
\def\ii{\'\i}
\begin{flushright}
ICN-UNAM-96-12\\
October 20th, 1996\\
\end{flushright}
\vskip 0.3truein
\centerline{{\bf\Large Boson Mapping in Field Theory}\footnote{
To be published in the ``Proceedings of the Conference on Nuclear Physics
at the Turn of the Millenium: Structure of Vacuum and Elementary Matter'',
South Africa, 1996}} 
\vskip 0.3truein
\centerline{P.O.Hess\footnote{HESS@ROXANNE.NUCLECU.UNAM.MX}, 
J.C.L\'opez\footnote{VIEYRA@ROXANNE.NUCLECU.UNAM.MX}, 
C.R.Stephens\footnote{STEPHENS@ROXANNE.NUCLECU.UNAM.MX}} 
\vskip 0.3truein
\centerline{\it Instituto de Ciencias Nucleares, UNAM,}
\centerline{\it Circuito Exterior, 
A.P. 70-543, 04510 M\'exico, D.F., Mexico}
\vskip 0.5truein

{
A boson mapping of pair field operators is presented. The mapping 
preserves all hermiticity properties and the Poisson bracket relations 
between fields and momenta. 
The most practical application of the boson mapping 
is to field theories which exhibit bound
states of pairs of fields.
As a concrete application we consider, in the low energy limit,  
the Wick-Cutkosky model with equal mass for the charged fields. }

\section*{I. Introduction} 

Boson mapping techniques play an important role in many areas of physics 
whenever the low lying states in energy 
are dominated by pair correlations of the fundamental particles. 
A typical example is the seniority model\cite{ring} where the dominant
contribution to the ground state of a nuclear system is given by
pairs of nucleons of total spin zero. In condensed matter physics
a famous example is superconductivity where pairs of
electrons are coupled via an electron-phonon interaction to form 
Cooper pairs. 

A pair 
of creation operators of two fundamental particles can be used to 
represent the creation of a composite bosonic field
if one can find a conjugate annihilation operator which
annihilates the field. A pair of annihilation
operators of the fundamental particles is not sufficient. However, the boson 
mapping\cite{ring,klein} allows one to find just such an 
operator.
%\footnote{One should not confuse the more general term boson mapping which may map
%a pair of fermions to bosons or bosons to bosons with the technique
%of bosonization which specifically applies to forming bosons from fermions.} 

It is natural to ask if such a mapping can be applied directly at the level of
field operators rather than annihilation and creation operators, 
the motivation being to map a Hamiltonian (Lagrangian) density
to an effective one which depends only on pair operators.
For example, if the Hamiltonian density of the gauge field of QCD 
can be mapped to one which depends only on composite operators  
representing
gluon pairs of total spin and colour zero, then one would end up 
with a new effective Hamiltonian density which depends only on a scalar field.
The simplification in structure is obvious. So, 
if at low energies the ground state is dominated by pair
correlations it is very probable that a boson mapping not only
simplifies the structure but also takes into account the physics. 
This is the case for QCD where perturbatively 
a pair of gluons of spin and colour zero is the lowest energy state.
Additionally, in the quark-anti\-quark sector we have ample experimental evidence
that the low energy physics is dominated by quark-anti\-quark pairs
of spin and colour zero (e.g., the pion) and by di-quarks.
However, up to now no explicit mapping at the level of field operators, 
depending on the coordinate $x=(\vec{x},t_o)$, where $t_o$ refers to a 
fixed time, has been given. All applications have been 
restricted to creation and annihilation operators\cite{ring,klein}.  
Recently, the boson mapping 
has been extended to pairs of coordinate operators 
$Q_{ai}=\frac{1}{\sqrt{2}}(b^\dagger_{ai} + b_{ai})$, 
where the indices $i$ and $a$ refer
to a set quantum numbers, and their derivatives 
$\frac{\partial}{\partial Q_{ai}}$\cite{hess}.
The main reason to consider such a  mapping is that the Hamiltonian densities
of field theories have a much simpler structure in terms of coordinate
operators $Q_{ai}$ which are the expansion coefficients of the fields
$\Phi_a(x)$ in terms of an orthonormal set of functions
$f_i(x)$, i.e.,
$\Phi_a(x)=\sum_i \frac{1}{\sqrt{\omega_i}} f_i(x) Q_{ai}$.

In this contribution we will show how to implement a boson mapping at the
level of classical pair fields, e.g., 
$\left\{ \frac{1}{\sqrt{N}}\sum_a \Phi_{a}(x) \Phi^a(x) \right\}=$  
$q(x)$ with $a$ denoting some quantum numbers,
and discuss some problems related to it. Although the mapping is presented
at the classical level it is relatively straightforward to check that it goes through
in the quantum case where Poisson brackets are replaced by commutators. 
As a specific example, we will
investigate the Wick-Cut\-kosky model\cite{wick} (WCM) which is known to have
bound states. In the 
WCM we integrate out the scalar field which mediates the interaction
between two charged fields. For equal masses of the latter 
and in the limit where the mass of the scalar field is much greater
than any relevant energy involved, 
we will obtain a model which, after boson mapping,
will be equivalent to a scalar field theory with only one scalar neutral
field. The classical equation of motion is that of an anharmonic
oscillator with an attractive fourth order term in the potential. 
Consequently, the classical system is unstable as 
$q \rightarrow \infty$
when all other classical degrees of freedoms are neglected. 
It is important to emphasize that in the full quantum theory this 
unphysical feature is expected to disappear.

We will not discuss here
the construction of the ground state and excited states of the WCM. 
This will be done in another publication  where we will compare results 
obtained from the boson mapping with those derived from a 
more fundamental renormalization group analysis of the model and 
investigate the limit of applicability of the boson mapping. 
We will also restrict our attention here to boson fields
as fundamental fields, the extension to fermion fields being straightforward.

\section*{II. A Boson Mapping for Fields:}  

Let us denote a general bosonic field by $\Phi_{ia}(x)$ with
$i=1,...,M$ and $a=1,...,N$. 
The conjugate momenta to these fields are denoted by
$\Pi^{ia}(x)$ and satisfy with $\Phi_{ia}(x)$ the 
Poisson bracket relation 
$\{ \Phi_{jb}(\vec{y},t_o), \Pi^{ia}(\vec{x},t_o) \} = $ $\delta_{jb}^{ia} 
\delta (\vec{x}-\vec{y})$.\footnote{The form of the Poisson bracket is given 
in the chapter about continous fields of the book written by 
H.Goldstein\cite{goldstein}. } We use co- and contravariant indices
in order to allow non-cartesian components. Associated to these fields and their
conjugate momenta we introduce pair fields ($q_{ij}(x)$) and
their conjugate momenta ($p^{ij}(x)$) satisfying the Poisson
bracket relation 
$\{ q_{nm}(\vec{y},t_o), p^{ij}(\vec{x},t_o) \} = (\delta_{nm}^{ij} +  
\delta_{nm}^{ji}) \delta (\vec{x}-\vec{y})$, where instead of the usual 
definition of the Poisson bracket we multiplied  
for convenience the right hand side by an extra factor of one half. 
(This comes first by noting that 
$\frac{\delta q_{nm}(x)}{\delta q_{ij}(y)} = $ 
$(\delta_{nm}^{ij}+\delta_{nm}^{ji}) \delta (x-y)$, which reflects the
non-normalized nature of $q_{ij}(x)$, and that we require the
above mentioned Poisson bracket relation between $q_{nm}(y)$ and
$p^{ij}(x)$ be satisfied in analogy to the usual boson 
pair relation\cite{klein}.)
The $q_{ij}(x)$ is called a pair field because it will be 
related to a pair of the fundamental fields $\Phi_{ia}(x)$.
It is still not normalized, however, as can be seen from the
Poisson bracket when $(nm)=(ij)$. This is also for convenience and can
later on be corrected by an appropriate normalization of the fields. 
Furthermore, the pair fields and their conjugate momenta are symmetric
with repect to interchange of their discrete indices
($q_{ij}(x)=q_{ji}(x)$, etc.).

We now propose a mapping from fundamental to composite fields 
such that the Poisson bracket
relations and hermiticity properties are conserved. Both conditions are
important: the former to ensure that the dynamics is preserved (for at least 
a physically relevant subspace of the space of states) and the latter to
ensure the invariance of matrix elements under the mapping.  
The explicit mapping is 
\begin{eqnarray}
\left\{ \frac{1}{\sqrt{N}}\sum_a \Phi_{ia}(x) \Phi_j^a(x) \right\} & = & 
q_{ij}(x)
\nonumber \\
\left\{ \frac{1}{\sqrt{N}}\sum_a \Pi^i_a(x) \Pi^{ja}(x) \right\} & = & 
\frac{1}{N} \sum_{k_1k_2} p^{ik_1}(x)q_{k_1k_2}(x)
p^{k_2j}(x)
\\
\left\{ \sum_a  \Phi_{ia}(x)\Pi^{ja}(x) \right\} & = &   
\sum_k q_{ik}(x) p^{kj}(x) 
\nonumber
\label{mapp}
\end{eqnarray}
where the curly bracket on the left hand side denotes the mapping,
i.e., the left hand side gives the operator in the original
space of states while the mapped expression gives the operator in the new
space of states, defined by the pair fields. 
In eq.\ref{mapp} both sides satisfy exactly the same Poisson bracket relations
as can be seen by direct calculation. The hermiticity properties are
obviously satisfied, i.e., the first two expressions on the left
are hermitian and so they are on the right. The last is anti-hermitian
and so is the mapped expression on the right. 
The $\frac{1}{\sqrt{N}}$ on the left hand side is the normalization
of the pair and has its origin in the coupling coefficient which
couples the field to a scalar with respect to the property 
characterized by the index $a=1,2,...,N$. 
For example, in the case of QCD this index
refers to colour and the above objects have colour zero. 
The mapping of eq.\ref{mapp} can easily be generalized to a quantized
theory, the only change being that the last operator has to be 
written symmetrically.

Note that a kinetic energy, quadratic in the momenta of the
original fields, is mapped to a kinetic energy quadratic in the pair fields 
but which now depends also on the field itself and not only on the conjugate
momenta. This corresponds to a ``mass" parameter which depends on the 
strength of the field. This situation appears quite often in physics.
As an example see ref.\cite{eisen} which describes 
the moment of inertia of a nucleus as a function of deformation. It can 
also be given the interpretation of working in a non-cartesian coordinate
system in the configuration space of the fields as will be seen in a less
general setting shortly. The appearance of the composite field explicitly 
in its own kinetic term indicates the high degree of non-linearity of the mapping, 
something that makes the explicit construction of the inverse mapping quite difficult. 

We have still to consider the mapping of 
expressions containing derivatives such as
$\sum_a (\nabla\Phi_{ia})\cdot (\nabla\Phi_j^a)$, which appear in
the original Lagrangian density. The problem here is that
the field derivative is a limit of differences of its value  
at two neighbouring points. Thus, the above product contains 
products of two fields at neighbouring points. The mapping in eq.\ref{mapp}
is only constructed for products of fields at the same point
(a generalization of it with different vectors $x$ and $y$ can also be given
but it leads to non-local field theories, which we try to avoid for the 
moment). For the case $M=1$, i.e. in the case where there are 
no external indices, the result is 

\begin{equation}
\left\{ \frac{1}{\sqrt{N}} \sum_a (\nabla\Phi_{a}(x)) \cdot
(\nabla\Phi^a(x)) \right\}
= \frac{(\nabla q(x)) \cdot (\nabla q(x))}{4q(x)}  \hspace*{0.5cm} .
\label{deriv}
\end{equation}
(For the general case it is more involved.)

We can also see this mapping of the derivative term, and indeed the
entire boson mapping itself, by considering a change of variables in the
measure of the functional integral. For instance, in
the case of a complex scalar field $\Phi (x)$ 
we change variable $\Phi (x)=\sqrt{q(x)} 
exp(i\theta(x))$. The volume element $d\Phi d\Phi^*$ changes to
$dq d\theta$, apart from a constant. The expression 
$(\partial_\mu \Phi )(\partial^\mu \Phi^*)$ 
becomes $\frac{(\partial_\mu q )(\partial^\mu q)}{4q}$
$+q(\partial_\mu \theta )(\partial^\mu \theta)$. 
For the spatial part
$(\nabla\Phi )\cdot (\nabla\Phi^*)$ the first term corresponds to 
the mapping, when restricted to pairs $q(x)$ only. 
For the part
$\mid \partial_o\Phi \mid^2$ we have to take the expression for the 
classical conjugate momentum $p(x)$ (given in the next section) 
whereupon we will arrive at the mapping
proposed in eq.\ref{mapp} for the square of the momenta $\Pi (x)$. 
Note, that in this example we have an extra contribution 
$q(\partial_\mu \theta )(\partial^\mu \theta)$. 
Genarically this type of term becomes 
important when the contribution of the fields
not coupled to pairs start to dominate. For that case the mapping
in eq.\ref{mapp} has to be generalized along the same line as done
in the examples of nuclear physics\cite{klein} when broken pairs
are introduced. The mapping in eq.\ref{mapp} is, therefore, restricted
(projected) to states of the space of states which are dominated
by the pairs $q(x)$. The simple change of variables, used in the above
example, is generalized in eq.\ref{mapp}.

In the future we intend to extend the mapping in eq.\ref{mapp} such
that pairs of fields at different space-time points are considered.
As mentioned above, this leads to non-local field 
theories, however, by using operator product expansion methods this yields
local field theories. The work is still in progress.

\section*{III. The Wick-Cutkosky model (WCM):} 

In the WCM\cite{wick} the Langragian, with equal masses of the 
charged fields, is given by 

\begin{eqnarray}
{\cal L}(x) & = & \frac{1}{2} \sum_{b=1}^2 
(\partial_\mu \Phi_b^\dagger (x))
(\partial^\mu \Phi^b(x)) + 
\frac{1}{2}( (\partial_\mu \phi(x)) (\partial^\mu \phi(x))
\nonumber \\
& & -\frac{M^2}{2} \sum_{b=1}^2 \Phi_b^\dagger (x) \Phi^b(x) 
-\frac{\mu^2}{2} \phi^2(x) \nonumber \\
& & -g(\sum_{b=1}^2 \Phi_b^\dagger (x) \Phi^b (x) ) \phi (x) 
\hspace*{0.5cm} .
\label{lagr}
\end{eqnarray}  
In eq.\ref{lagr} the charged field is described by a complex field which
in terms of real and imaginary parts implies that the sum over $a$ 
involves four real fields. In this case it manifestly has the structure of 
eq.\ref{mapp}\footnote{In fact the mapping in eq.\ref{mapp} is the same for 
complex fields in which one of the fields has to be taken as the complex
conjugate and the value $N$ is twice the number of complex fields.}.

Using functional integration with Euclidian measure we can integrate 
over the real field $\phi (x)$ using gaussian 
integration. 
We obtain finally for an effective Lagrangian
density

\begin{eqnarray}
{\cal L}_{eff}(x) & = & 
\frac{1}{2} \sum_{b=1}^2 
\partial_o\Phi_b^\dagger (x) \partial_o\Phi^b(x) - 
\frac{1}{2} \sum_{b=1}^2 
(\nabla \Phi_b^\dagger (x)) \cdot
(\nabla \Phi^b(x)) + 
\nonumber \\
& & -\frac{M^2}{2} \sum_{b=1}^2 \Phi_b^\dagger (x) \Phi^b(x) 
\nonumber \\
& & + \frac{g^2}{2} (\sum_{b=1}^2 \Phi_b^\dagger (x) \Phi^b(x))
\int d^4y \Delta_F(x - y) 
( \sum_{c=1}^2 \Phi_c^\dagger (y) \Phi^c(y) )
\label{leff}
\end{eqnarray} 
where $\Delta_F(x - y)$ is the Feynman propagator
for scalar fields. 

In the limit of $\mu^2$ much greater than the momenta involved
which corresponds 
for a given ultraviolet cutoff 
$\Lambda$ to $(\frac{\mu}{\Lambda})^2>>1$, i.e. that
the momenta involved are within the
low energy limit, the interaction part of the 
effective Lagrangian  density in eq.\ref{leff}
acquires the form

\begin{eqnarray}
\frac{g^2}{2\mu^2} 
(\sum_{b=1}^2 \Phi_b^\dagger (x) \Phi^b(x))^2
\hspace*{0.5cm} .
\label{intereff}
\end{eqnarray}

We now express the complex fields $\Phi_b(x)$ in terms of their real 
and imaginary part, i.e., $\Phi_b(x)=\Phi_{1,b}(x)+\Phi_{2,b}(x)$ which
yields four different fields $\Phi_a(x)$ with a linearized
index $a=1,...,4$ ($a=1$: $(1,b)$; $a=2$: $(2,b)$; etc.). Then we change  
to the Hamiltonian formulation because the mapping involves the conjugate 
momenta. This yields for the kinetic energy the form
$\frac{1}{2} \sum_{a=1}^4 \Pi_a (x) \Pi^a(x)$ and the potential part
in eq.\ref{leff}
(including the term which involves the square of the gradient of the fields)
changes its sign.
With the  mapping of eq.\ref{mapp} (with $i,j=1$)
and the normalization of the pair fields $q(x)$ by multiplying them and
their conjugate momenta by $\frac{1}{\sqrt{2}}$ (denoting them afterwards
with the same letters) this Hamiltonian density is mapped to

\begin{eqnarray}
{\cal H}_{bos}(x) & = & 
\frac{1}{\sqrt{2}}p(x) q(x) p(x) +\sqrt{2}M^2 q(x)  
-\frac{4g^2}{\mu^2}q^2(x)  \nonumber \\ 
& & +\frac{\sqrt{2}}{4}\frac{(\nabla q(x)) \cdot (\nabla q(x))}{q(x)} 
\hspace*{0.5cm} .
\label{lbos}
\end{eqnarray}
The kinetic energy of the Hamiltonian density contains a dependence 
on the field $q(x)$ and the potential has a term $\sim q(x)$ which can
be interpreted as a magnetic type of interaction. 
 From eq.\ref{lbos} the momentum as a function of $q(x)$ can be deduced:

\begin{equation}
p(x) = \frac{1}{\sqrt{2}q(x)}\frac{\partial q(x)}{\partial t}
\label{p(x)}
\end{equation}

The equation of motion is derived from the new Hamiltonian density 
and is of the form

\begin{eqnarray}
\Box q(x)-\frac{1}{2q(x)}(\nabla^\mu q(x))\cdot (\nabla^\mu q(x))
\nonumber \\
+ 2M^2q(x)-\frac{8\sqrt{2}g^2}{\mu^2}q^2(x) = 0  
\hspace*{0.5cm} .
\label{eqmotion1}
\end{eqnarray}
The second term is a connection term associated with, as mentioned
previously, the fact that we are using non-cartesian coordinates.
Making the substitution  $q(x)=\chi^2(x)$ we arrive at the equation

\begin{equation}
\Box \chi (x) 
+ M^2\chi (x)-\frac{4\sqrt{2}g^2}{\mu^2}\chi^3(x) = 0 
\hspace*{0.5cm} .
\label{eqmotion2}
\end{equation}
This is the classical equation of motion 
of an anharmonic oscillator where the
anharmonic term in the potential is attractive. 
This term destabilizes the system for large coupling constant $g$
and produces a transition to a state with infinite expectation
value $<q>$. It, therefore, indicates that the system is unstable under
the formation of a $q$-condensate. Higher, repulsive terms, not included
in the Hamiltonian density, should make it stable.

The approximation we have made in the above 
is restricting the space of states
to functionals which depend on the pair fields only. It remains to be seen
that at low energy this assumption is justified, i.e., that the
lowest lying states are dominated by bound states which are comprised
of a coupling of two fields. This is under investigation\cite{stephens}.

\section*{IV. Conclusion} 

In this contribution we proposed a mapping of fundamental 
pair fields and their conjugate momenta to new composite fields
and their corresponding conjugate momenta.
We restricted attention to the case of classical fields, 
with the proviso that the mapping can be 
directly extended to a quantized picture without undue difficulty. 
The mapping will be particularly useful whenever 
pair correlations dominate the low energy
structure, as is the case for QCD. Only boson fields as fundamental
fields were considered, the extension to fermionic fields being direct.
(There is some relation of the boson mapping, as presented here on the
operator level, to the formulation with Feynman path 
integrals\cite{scholtz}. We thank F.J.W.Hahne for pointing this out during
the conference.)

The mapping was applied to the Wick-Cutkosky model, where in the
limit of low energy and equal masses for the charged fields we obtained
a field theory equivalent to that of a neutral self-inter\-acting
scalar field. 
The classical equation of motion was that 
of an anharmonic oscillator with an attractive 
anharmonic term of fourth order 
in the potential. As a result, the system 
is unstable to small fluctuations.

What still has to be shown in a fundamental (basic) calculation
is that at low energies the Wick-Cutkosky
model exhibits the pair structure used. Also under investigation is how
far the boson model gives correct results compared to the exact
description. This is important in order to probe the range of 
application of the boson model and deduce from there possible 
applications to QCD.

\end{document}